\documentclass{article}

\usepackage{arxiv}

\usepackage[utf8]{inputenc} % allow utf-8 input
\usepackage[T1]{fontenc}    % use 8-bit T1 fonts
\usepackage{hyperref}       % hyperlinks
\usepackage{url}            % simple URL typesetting
\usepackage{booktabs}       % professional-quality tables
\usepackage{amsfonts}       % blackboard math symbols
\usepackage{nicefrac}       % compact symbols for 1/2, etc.
\usepackage{microtype}      % microtypography
\usepackage{lipsum}		% Can be removed after putting your text content
\usepackage{graphicx}
\usepackage{natbib}
\usepackage{doi}
\usepackage{amsmath}

\title{Latent State Space Extension for interpretable hybrid mechanistic models}

\newif\ifuniqueAffiliation
\uniqueAffiliationtrue
\ifuniqueAffiliation
\author{ \href{https://orcid.org/0009-0009-1196-305X}{\includegraphics[scale=0.06]{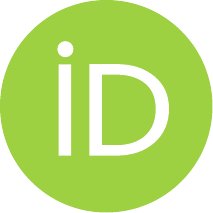}\hspace{1mm}Judit Aizpuru},        \href{https://orcid.org/0000-0002-1493-0319}{\includegraphics[scale=0.06]{orcid.pdf}\hspace{1mm}Maxim Borisyak},
    \href{https://orcid.org/0000-0002-1214-9713}{\includegraphics[scale=0.06]{orcid.pdf}\hspace{1mm}Peter Neubauer},
    \href{https://orcid.org/0000-0001-9461-4414}{\includegraphics[scale=0.06]{orcid.pdf}\hspace{1mm}M. Nicolas Cruz Bournazou}\\
	Technische Universität Berlin \\
    Chair of Bioprocess Engineering \\
    Berlin, Germany \\
	\texttt{j.aizpuru@campus.tu-berlin.de}
}
\renewcommand{\shorttitle}{\textit{Latent State Space Extension for interpretable hybrid mechanistic models}}

\hypersetup{
pdftitle={aizpuru_hybrid},
pdfsubject={q-bio.NC, q-bio.QM},
pdfauthor={Judit Aizpuru},
pdfkeywords={mechanistic model, hybrid model, model discovery, bioprocess control, parameter dynamics},
}

\begin{document}
\maketitle

\begin{abstract}
Mechanistic growth models play a major role in bioprocess engineering, design, and control. Their reasonable predictive power and their high level of interpretability make them an essential tool for computer aided engineering methods. Additionally, since they contain knowledge about cell physiology, the parameter estimates provide meaningful insights into the metabolism of the microorganism under study. However, the assumption of time invariance of the model parameters is often violated in real experiments, limiting their capacity to fully explain the observed dynamics. In this work, we propose a framework for identifying such violations and producing insights into misspecified mechanisms. The framework achieves this by allowing kinetic and process parameters to vary in time. We demonstrate the framework’s capabilities by fitting a hybrid model based on a simple mechanistic growth model for E. \emph{coli} with data generated in-silico by a much more complex one, and identifying missing kinetics.
\end{abstract}

% keywords can be removed

\section{Introduction}
Mechanistic models are widely applied in bioprocess development, from offline design of experiments to controlling and re-designing experiments in an online manner \cite{kim2022model}. They are typically composed of differential equations describing the kinetics of the uptake and production of different metabolites and biomass, and mass transport and conservation laws. They are typically derived under assumptions of time-invariant parameters and contain the dynamics of just some of the macroscopic species. This makes them simpler and easier to fit with a limited number of experimental measurements, but ignores the very complex metabolic adaptation mechanisms cells are provided with to cope with the cultivation environment, and the possible effects of unmodelled dynamics. This simplification evidently limits their predictive performance since these assumptions are violated in most cultivating conditions, making long term predictions challenging.

In the literature, the problem has been addressed by using hybrid models. These types of models try to use the knowledge that is embedded in the mechanistic models complementing them with data-driven parts. In the literature one can encounter hybrid models where a data-driven component is coupled to the right-hand side of the mechanistic model, for example, by addition of a neural differential equation network to capture missing dynamics \cite{quaghebeur2022hybrid}, or by multiplication of a network to the kinetic reaction rates \cite{oliveira2004combining}.

In this work, we propose a framework for hybrid modelling inspired by \cite{rangapuram2018deep}, where the authors propose an explainable model based on a linear differential equation system where the parameters can vary in time being the outputs of a recurrent neural network. Our framework in contrast, introduces a linear latent state space model that drives the mechanistic part. Making the mechanistic model parameters linear combinations of these latent states allows them to vary in time, keeping the original interpretation of the model while being flexible enough to express arbitrary adaptation behaviours. We utilise this flexibility to gain important insights into the metabolic activity and the response of the cells to environmental changes. This allows us to detect any deviations from the assumptions and identify possible misspecified mechanisms in the mechanistic model. We demonstrate the framework’s capabilities by fitting a hybrid model based on a simple mechanistic growth model for E. \emph{coli} with data generated in-silico by a much more complex one, and detecting the missing kinetics from the base model.
 
\section{Hybrid Model}

Consider a mechanistic model \emph{f}, to which we refer as \emph{the
base model} everywhere below:

\begin{equation}
\label{eq:base}
    \frac{dx}{dt} = f\left( x,\theta \right)    
\end{equation}

where \emph{x} is a vector of states, \emph{$\theta$} denotes the vector of
kinetic and process parameters of the model. As an example, we use the E.
\emph{coli} growth model from \cite{anane2017modelling} without the acetate dynamics,
thus, composed of three states: biomass (X), glucose (S), and dissolved
oxygen (DOT). The uptake of substrate will be given by a Monod term, and
oxygen is modelled as the difference of oxygen transfer and uptake rate,
the latter obtained by mass balancing with glucose.

The proposed hybrid model extends the state space of the base model with
latent variables \emph{z}, whose dynamics are linear with respect to the
mechanistic states \emph{x} and themselves. Linear combinations of these
states describe the time evolution of the kinetic and process parameters
\emph{$\theta$} of the base model as:

\begin{equation}
\begin{aligned}
\label{eq:latent_extension}
    \frac{dz}{dt} = Az + Bx \\
    \theta\left( z \right) = Wz + \theta_{0}
\end{aligned}
\end{equation}

Where \(x \in R^{n}\) and \(z \in R^{m}\) are the base model's state
space and states of the model's extension. \(\theta_{0} \in R^{p}\) are
the values for the base model's parameters when no latent dynamics are
present. Finally, \emph{A, B} and \emph{W} are parameters of the latent
extension, represented as matrices with dimensions \(m \times m\),
\(m \times n\), and \(p \times m\) respectively.

The proposed hybrid model allows varying parameters of the base model in
time, and in addition, it reduces to the base model when the latent
states are constant in time. It is worth noting that the latent
extension is a universal approximator, i.e., it is capable of expressing
arbitrary smooth functions (given that the latent dimensionality is
large enough). The main idea behind the proposed method is to detect
deviations between the process and the base model by the presence of
parameter dynamics.

\section{Method}

In order to prevent the hybrid model from producing unnecessary
parameter dynamics (for example, due to overfitting to measurement
noise), and ensure that non-constant dynamics are present only to
compensate for the bias in the base model, we introduce a regularisation
term to the loss function:

\begin{equation}
\label{eq:loss}
    loss = - \sum_{i = 1}^{n}{\sum_{j = 1}^{N_{i}}\frac{\left( x_{j}^{i} - y_{j}^{i} \right)^{2}}{\sigma_{i}^{2}}} + \lambda\left( \sum_{i = 1}^{n}{\sum_{j = 1}^{n}a_{i,j}^{2}} + \sum_{i = 1}^{m}{\sum_{j = 1}^{n}b_{i,j}^{2}} + \sum_{i = 1}^{p}{\sum_{j = 1}^{m}w_{i,j}^{2}} \right)
\end{equation}

Where $\lambda$ controls the strength of the regularisation.
Additionally, we restrict the norm of the initial latent state.

It is worth noting three important properties of the model. On the one
hand, the proposed model is a strict extension of the base model, i.e.,
it is able to produce the same dynamics as those to the base model when
setting matrices A, B and W to null ones. On the other hand, absolute
values of matrix elements are bounded from above by
$\sqrt{\frac{L_{0}}{\lambda}}$, where $L_{0}$ is the loss of the
base model, and therefore, the matrix norms are bounded as well. Since
x(t) is bounded by the loss function, and latent dynamics are bounded by
the norm of matrices and the norm of the initial latent state, the
parameter dynamics are restricted and dependent on $L_{0}$. Most importantly, any non-negligible parameter dynamics would imply
mismatch between the base model and the data. As the regularization term
promotes minimal change in the hybrid model's parameters, the dynamics
are likely to be related to misspecified kinetics of the base model,
thus, offering a valuable insight into these missing mechanisms.

To select an appropriate value for the regularization strength and
prevent overfitting, we suggest selecting \(\lambda\) experimentally by
fitting the hybrid model to data generated from the base model, with
noise, and gradually increasing \(\lambda\) until estimated parameter
dynamics become negligible.

\section{In-silico experiments}

In order to demonstrate the capabilities of the proposed modeling
framework, we fit the base and extended model to data generated from a
more complex one. This allows us to clearly demonstrate the proposed
framework as the differences between the models are known analytically.
We select a test scenario that has the usual characteristics of a real
experiment. As ground-truth model, we used a slightly modified version
of the E. \emph{coli} growth model by (Anane et al. 2017). The model
state contains biomass, glucose, acetate (measured at-line), and
dissolved oxygen (measured online). 10 different datasets were generated
using different sets of parameters and initial conditions. After the
batch phase, bolus feeding is applied with constant feed volumes every
12 minutes. Finally, normally distributed noise is added to the
observations, using realistic magnitudes. Both models, the base and the
hybrid ones, were fitted to each dataset, using a 4-dimensional latent
space for the hybrid model.

For simulation, a fixed step Euler integrator was used. With a little
modification in the model so that the stiffness is relaxed, this
integrator allows fast simulation without much difference in the
solution to a more complex multi-step method. For the optimization, the proposed loss function in Equation \ref{eq:loss} has been
minimized using an Adam optimizer. To avoid bounded optimization, all
parameters of the base model and the corresponding time-varying
parameters of the hybrid model are transformed with inverse error
function (scaled and shifted by the corresponding ranges) rendering them
unbounded. Additionally, to avoid local minima, we use a multi-start
procedure with 50 random initial guesses.

\section{Results}
\subsection{Interpolation and extrapolation error}
For evaluating the performance of the hybrid model in comparison to the
base model, we consider two metrics: interpolation and extrapolation
errors. For the calculation of the former, 100 points from the ground
truth simulation were sampled randomly. The sum of squared standard
errors (SSE) was calculated for each of the models for each dataset. For
the calculation of the extrapolation error, 50 observations equally
spaced in time within the future 2h were sampled from the simulation.
For fitting the hybrid model, we used $\lambda=0.01$. Figure \ref{fig:violin} shows
the distributions and quartiles of these errors, and an overall better
performance of the hybrid model for both interpolation and extrapolation
tasks. This clearly shows that the hybrid model is expressive enough to
approximate the effects of the missing kinetics. Comparing losses to the
level expected due to noise, we confirm that: the base model is indeed
biased, hybrid model is not overfitted, and, thus, lambda is selected
appropriately.

\begin{figure}[h]
\centering
\includegraphics[width=2.5904in,height=1.94298in]{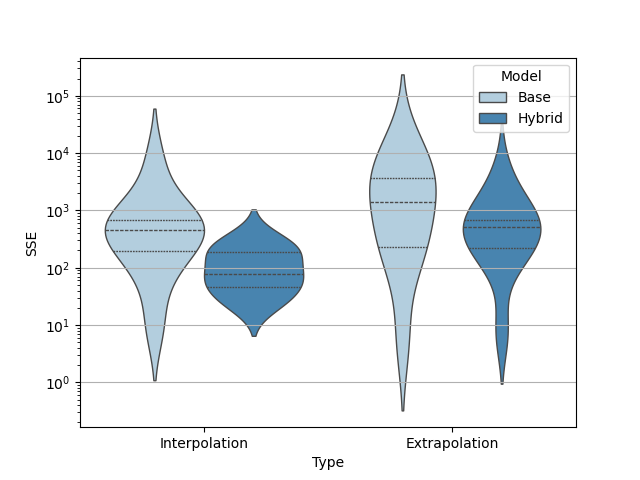}
\caption{Violin plots showing the distributions of the validation errors for each of the datasets and each of the models. The quartiles of the distributions are highlighted with dashed lines.}
\label{fig:violin}
\end{figure}

\subsection{Comparison of two solutions with different levels of acetate accumulation}

The base model misses the mechanisms associated with acetate production
and consumption. Therefore, in the absence of this metabolite, one
should expect both models' predictions to match and the base model to
yield a good fit. On the contrary, as soon as a noticeable concentration
of this substance is present in the cultivation, one would expect to it
to fail, as the kinetic rate for glucose consumption and growth rate are
affected by it. We have chosen two scenarios where this effect takes
place.

In Figure 2, the low-acetate scenario (the top graph), it can be seen
how both models achieve a good fit. At the same time, we observe that
parameter dynamics of the hybrid model remain nearly constant (Figure 3,
the top graph). The mismatch between the values of the two models might be
due to the usual identifiability issues, however, as can be seen from
the graph, predictions are nearly equivalent. The second case corresponds to an
accumulation of around 0.4g/l~ (Figure 2, the bottom graph). It is
noticeable how in the second scenario the base model fails to properly
fit the biomass dynamics at the end of the process, and the upper level
of oxygen is not well captured during the fed-batch either. The hybrid
model is able to compensate for the missing dynamics and improve the fit
of both of these quantities. The parameter dynamics (Figure 3, the
bottom graph) show how affinity and yield on glucose (\(K_{S}\) and
\(Y_{\text{XS}}\)) decrease, while the maintenance uptake grows
(\(q_{m}\)), suggesting there is a larger substrate uptake coinciding
with decreased growth of biomass, which happens when acetate is present
in the culture. At the same time, the parameters related to oxygen
suggest that the saturation of oxygen does not reach the physical
maximum (\(\text{DO}T^{*}\)), effect encountered when the acetate has
not been re-cycled. In addition, the increased volumetric oxygen
transfer rate (\(k_{L}a\)), might be related to the extra oxygen cells
use for metabolizing acetate.

\begin{figure}[h]
\centering
\includegraphics[width=4.7in,height=1.5in]{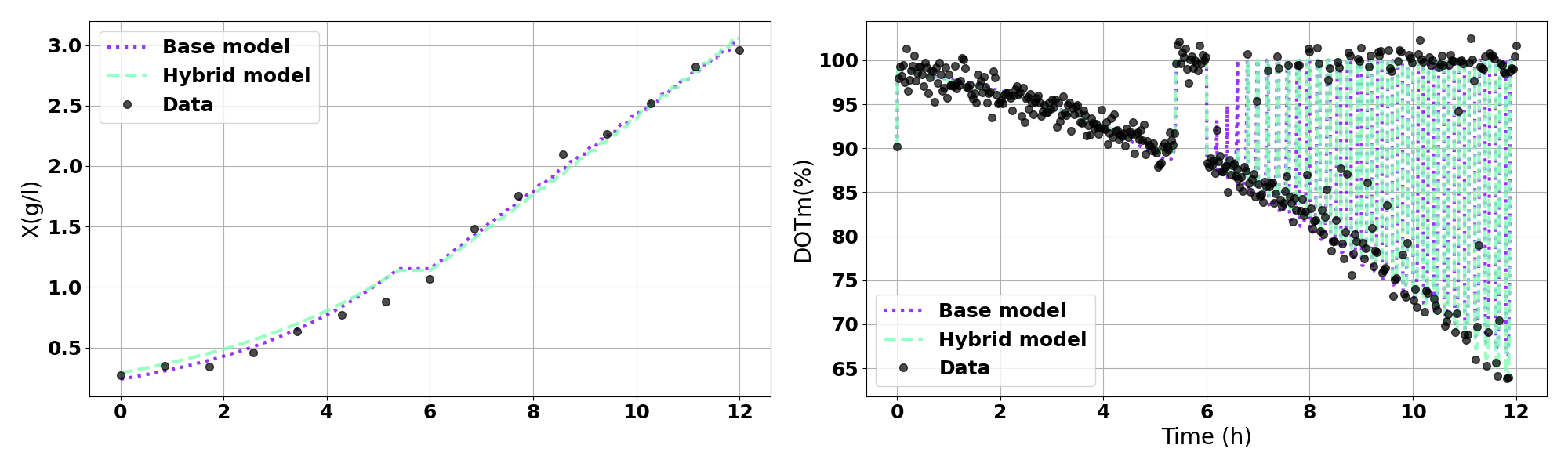}
\includegraphics[width=4.7in,height=1.5in]{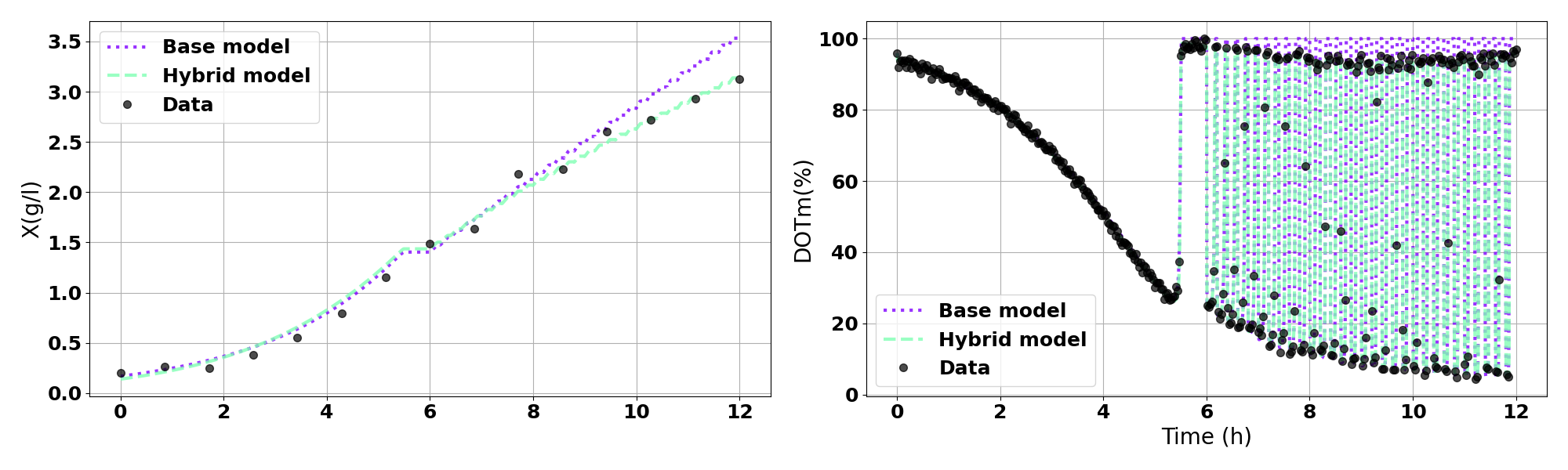}
\caption{Time-series of the solutions for two optimization scenarios,
low acetate (above) and high (below), for the base and the hybrid model,
together with the in-silico experimental
data.}
\label{fig:violin}
\end{figure}

\begin{figure}[p]
\centering
\includegraphics[width=\linewidth]{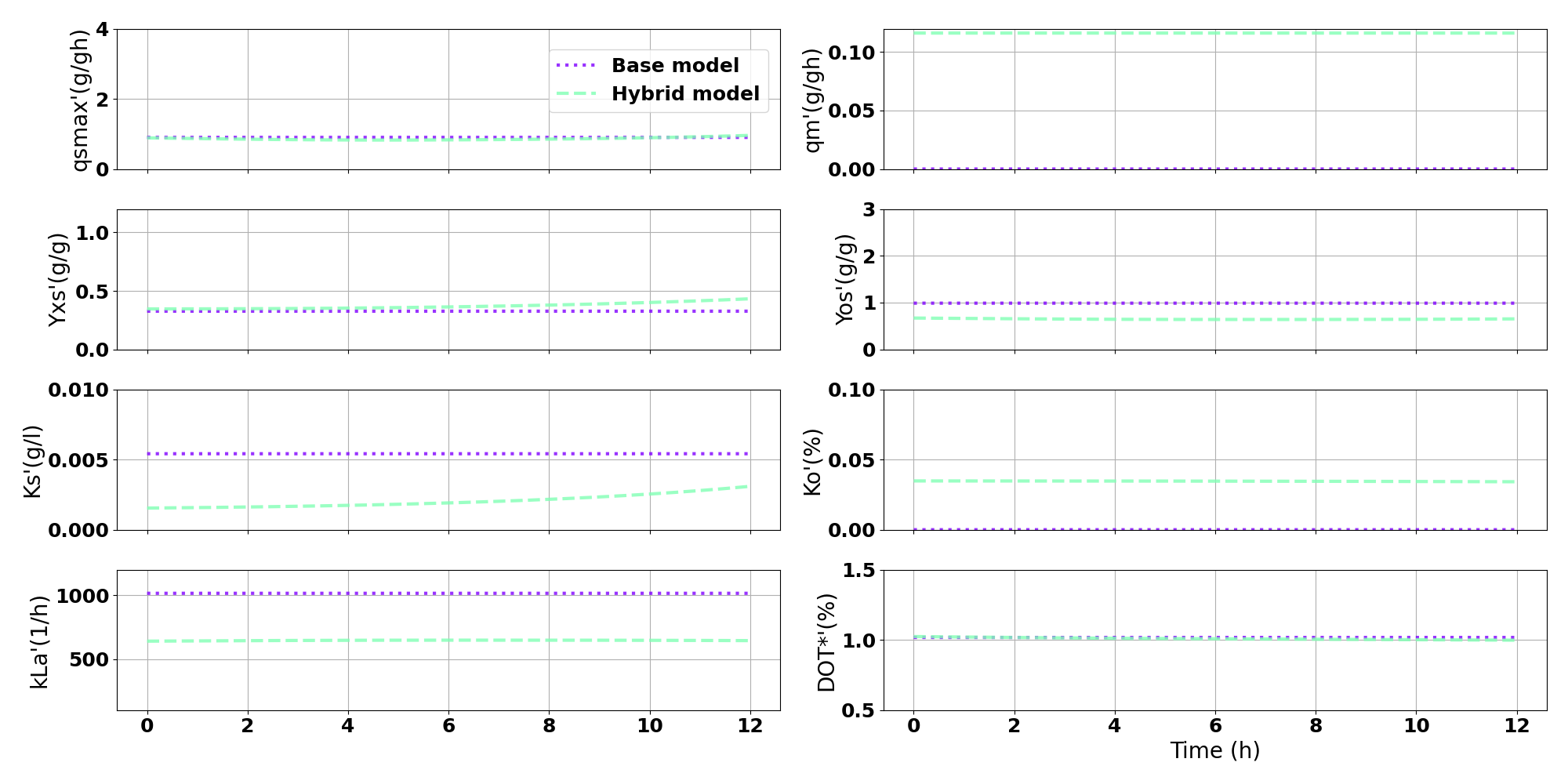}
\includegraphics[width=\linewidth]{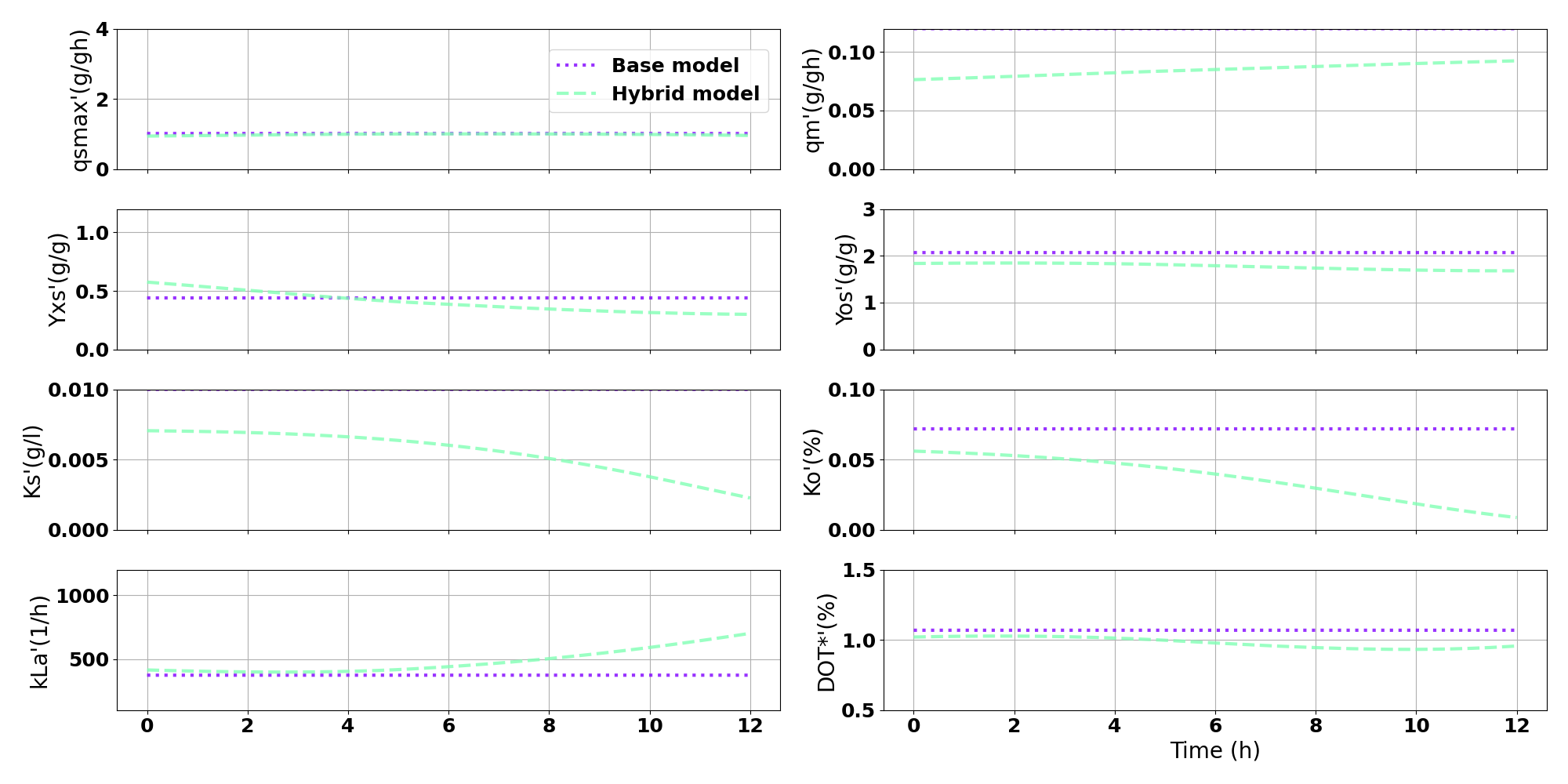}
\caption{Parameter dynamics predicted by the hybrid model comparing to
the static estimates from base model. For the low acetate scenario
(above) and high (below).}
\label{fig:violin}
\end{figure}

Overall, we observe that parameters that demonstrate significant dynamics are all related to the missing kinetics, namely the inhibition of substrate uptake due to acetate and additional oxygen consumption due to incomplete acetate recycling during the feed pulses. It is important to note that the parameter dynamics do not necessarily approximate the actual behaviour of the missing mechanisms, however, they still point to the misspecified parts of the base model.

\section{Conclusion}
In this work, an explainable and novel hybrid model has been presented as a framework for identifying and analyzing misspecifications of macro-kinetic models. The model demonstrates high predictive capacity as shown by our experiments. The proposed fitting procedure allows detecting any deviations from the observed process, as the hybrid model diverges from the base model only when the latter is not sufficient for a proper fit. It serves as a tool for verifying the applicability of existing models, and potentially developing new ones.

\section*{Acknowledgements}
We gratefully acknowledge the financial support of the German Federal Ministry of
Education and Research (01DD20002A – KIWI biolab).

%\newpage
\bibliographystyle{unsrtnat}
\bibliography{references}  

%%% Uncomment this line and comment out the ``thebibliography'' section below to use the external .bib file (using bibtex) .





% \end{thebibliography}

\end{document}